\documentclass[twocolumn,prc,amsmath,amssymb]{revtex4}
\usepackage{graphicx}
\usepackage{array}
\usepackage{dcolumn}
\usepackage{bm}
\newcolumntype{.}{D{.}{.}{-1}}
\begin{document}

\title{Precision Measurement of the $^{29}$Si, $^{33}$S, and $^{36}$Cl
  Binding Energies}

\author{M. S. Dewey}
\email{maynard.dewey@nist.gov}
\author{E. G. Kessler, Jr.\@}
\email{ernest.kessler@nist.gov}
\author{R.D. Deslattes}
\thanks{Deceased}
\affiliation{%
  National Institute of Standards and Technology\\
  Gaithersburg, MD 20899}%

\author{H.G. B{\"{o}}rner}
\author{M. Jentschel}
\author{C. Doll}
\author{P. Mutti}
\affiliation{Institut Laue-Langevin, F-38042 Grenoble CEDEX, France}

\date{\today}

\begin{abstract}
  The binding energies of $^{29}$Si, $^{33}$S, and $^{36}$Cl have been
  measured with a relative uncertainty \hbox{$< 0.59 \times 10^{-6}$}
  using a flat-crystal spectrometer.  The unique features of these
  measurements are 1) nearly perfect crystals whose lattice spacing is
  known in meters, 2) a highly precise angle scale that is derived
  from first principles, and 3) a gamma-ray measurement facility that
  is coupled to a high flux reactor with near-core source capability.
  The binding energy is obtained by measuring all gamma-rays in a
  cascade scheme connecting the capture and ground states.  The
  measurements require the extension of precision flat-crystal
  diffraction techniques to the 5 to 6 MeV energy region, a
  significant precision measurement challenge.  The binding energies
  determined from these gamma-ray measurements are consistent with
  recent highly accurate atomic mass measurements within a relative
  uncertainty of $4.3 \times 10^{-7}$.  The gamma-ray measurement
  uncertainties are the dominant contributors to the uncertainty of
  this consistency test.  The measured gamma-ray energies are in
  agreement with earlier precision gamma-ray measurements.
\end{abstract}

\maketitle

\section{Introduction}
\label{sec:intro}

Nuclear binding energy measurements are of interest because they are
accurately related to atomic mass measurements. This relationship
provides a means to check the results in one precision measurement
field against related results obtained in another precision
measurement field. Because the experimental techniques used in the two
fields are very different, this check has the potential to reveal
systematic errors associated with either measurement.

The synergism between binding energy and atomic mass measurements can
be demonstrated by considering a typical neutron capture reaction
$n+{}^{A}\text{X} \rightarrow {}^{A+1}\text{X}+\gamma$'s which leads
to the following equation involving atomic masses and the binding
energy,
\begin{equation}
  \label{eq:ambe}
  m(n) + m(^{A}\text{X}) = m(^{A+1}\text{X}) + S_{n}\,.
\end{equation}
Atomic masses $m$ are measured in atomic mass units while the binding
energy of ${}^{A+1}\text{X}$, $S_{n}$, is obtained from gamma-ray wavelengths
measured in meters.  The binding energy in meters can be converted to
atomic mass units using the molar Planck constant, $N_{A}h$, divided
by the speed of light, $c$ \cite{desl79}.  This combination of
constants is known with a relative uncertainty of $6.7 \times
10^{-9}$, an accuracy which does not limit the test implied by
Eq.~(\ref{eq:ambe}), given the presently available accuracy in atomic
mass and binding energy measurements \cite{mohr02}.

New precision measurements of the $^{29}$Si, $^{33}$S, and $^{36}$Cl
binding energies have been made using a flat crystal spectrometer.
This spectrometer measures the wavelengths of the gamma-ray photons
using crystals whose lattice spacings are known in meters and an angle
scale that is derived from first principles.  Thus, the measured
wavelengths are on a scale consistent with optical wavelengths and the
SI definition of the meter.  The binding energies of these three
nuclei are in the 8.5 to 8.6 MeV range and are obtained by measuring
lower energy lines that form a cascade scheme connecting the capture
and ground states.  For all three nuclei, the cascade scheme with the
most intense transitions includes a gamma-ray with energy $>
4.9$\,MeV. Such high energies present a significant measurement
challenge for gamma-ray spectroscopy because the Bragg angles are $<
0.1^{\circ}$ and the diffracted intensity is rather small (a few
s$^{-1}$ or less).

\section{The $^{29}$Si, $^{33}$S, and $^{36}$Cl Decay Schemes}
\label{sec:decay}

\begin{figure}
  \includegraphics[width=3.4in]{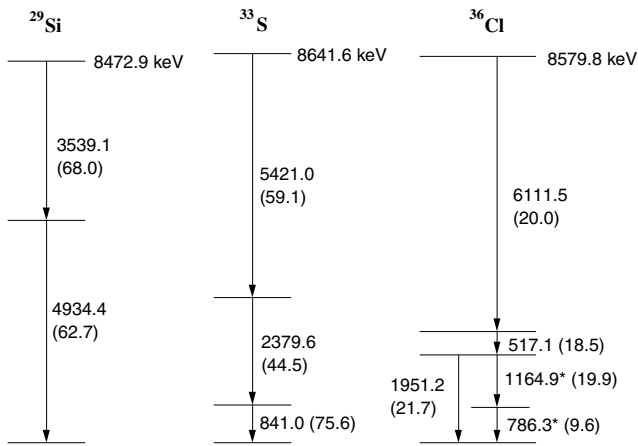}
  \caption{\label{fig:decay}Partial decay schemes for $^{29}$Si,
    $^{33}$S, $^{36}$Cl showing the transitions that were measured in
    this study. The numbers in parentheses are the number of gammas
    per 100 neutron captures.  The transitions marked with asterisks
    in Cl were not used to deduce binding energies.}
\end{figure}

In Fig.~\ref{fig:decay} we show partial decay schemes for $^{29}$Si,
$^{33}$S, and $^{36}$Cl containing the transitions that were measured
in these binding energy determinations.  The values in parentheses are
the number of gamma-rays emitted per 100 captures \cite{lone81}. The
reactions, the thermal neutron capture cross sections, and the nominal
energies of the measured gamma-rays are listed in
Table~\ref{tab:reactions} \cite{fire96}. Because $^{35}$Cl has the
largest thermal neutron capture cross section of any light nuclei, the
$^{36}$Cl binding energy was measured first.  The experience gained in
increasing the crystal reflectivity and lowering the background during
the $^{36}$Cl measurement proved to be very valuable in the
measurement of the weaker $^{29}$Si and $^{33}$S transition energies.
In addition, a larger than expected dependence of the angle
calibration on the environment was noted during the Cl measurements
which led to more frequent angle calibrations during the $^{29}$Si and
$^{33}$S measurements.  The $^{29}$Si measurement is particularly
difficult because the 4934 keV line is emitted while the nucleus is
recoiling following the emission of the 3539 keV line.  Thus, the 4934
keV profile is significantly Doppler broadened which decreases the
accuracy with which the Bragg angle can be determined.  In each of
these nuclei there are other less intense transitions that connect the
capture and ground states.  However, the measurement uncertainty of
these weaker lines will be so large that their contribution to the
binding energy determination will be small.  As the stability of the
spectrometer improves and techniques to measure weaker lines are
developed, these weaker lines may be used for future binding energy
determinations.

\begin{table*}
  \caption{\label{tab:reactions}Reactions, cross-sections, and
  nominal energies associated with the binding energy determinations.}
\begin{ruledtabular}
\begin{tabular}{l..cc}
\multicolumn{1}{c}{Nuclide} & \multicolumn{1}{c}{Reaction} &
        \multicolumn{1}{c}{Cross-section} & Nominal &
        \multicolumn{1}{c}{Energies measured} \\
        & & \multicolumn{1}{c}{($\times10^{-28}$ m$^{2}$)}  & binding
        energy & \multicolumn{1}{c}{(keV)} \\
        & & & (keV) & \\
\hline
$^{29}$Si & \multicolumn{1}{c}{$n + {}^{28}\text{Si} \rightarrow
        {}^{29}\text{Si} + \gamma$} & 0.177 & 8473 & 3539, 4934 \\
$^{33}$S & \multicolumn{1}{c}{$n + {}^{32}\text{S} \rightarrow
        {}^{33}\text{S} + \gamma$} & 0.53 & 8642 & 841, 2380, 5421 \\
$^{36}$Cl & \multicolumn{1}{c}{$n + {}^{35}\text{Cl} \rightarrow
        {}^{36}\text{Cl} + \gamma$} & 43.6 & 8580 & 517, 786, 1951, 6112 \\
\end{tabular}
\end{ruledtabular}
\end{table*}

\section{Experiment}
\label{sec:exper}

The measurements were made at the Institut Laue-Langevin (ILL) using
the GAMS4 flat crystal spectrometer.  A detailed description of this
facility is available to the interested reader in
Ref.~\onlinecite{kess01}.  The discussion of the experiment given here
will be limited to those aspects that are peculiar to these
measurements.  This spectrometer is located on the reactor floor at
the exit of a through-tube that has facilities to transport and hold
sources next to the reactor core.  Five beamtime allocations were
devoted to these binding energy measurements.  Chlorine was measured
in February 1997 and September 1997.  Sulfur was measured in
September--October 1998 and April--May 1999.  Silicon was measured in
October--November 2000.  The first measurement cycle (February 1997)
served as a proof of the capability of the facility to determine
binding energies.  In retrospect, it was necessary to exclude all data
taken during this cycle from the final results because the
experimental conditions were not sufficiently controlled.

The stability of the angle calibration of the spectrometer has been a
particularly troublesome experimental problem.  Throughout the
extended period of the measurements reported here, the angle
calibration of the spectrometer has been measured many times, both
during and between the binding energy measurements.  However, at the
start of these measurements we were not aware of the dependence of the
angle calibration on the humidity and did not perform calibrations as
frequently.  From 1998 forward, more frequent calibrations have
provided a means to obtain more accurate angle calibrations.
Following considerable data analyses aimed at mining the angle
calibrations and gamma-ray measurements for maximum information, we
have reached the following conclusions concerning the angle
calibration: a relative uncertainty due to the angle calibration of
$\approx 0.4 \times 10^{-6}$ should be included in each measurement
period.  More details concerning the angle calibration are given in
Sec.~\ref{sec:angcal}.

At the start of the experiment, it was our intention to measure the
786 keV and 1165 keV lines in Cl to provide a cascade cross-over
verification of the 1951 keV line.  Because of beamtime restrictions
and angle calibration difficulties, this cascade cross-over
verification was not realized.  However, a very limited data set for
the 786 keV line was recorded and used to provide a value for the
wavelength of this transition.

\subsection{Sources}
\label{subsec:sourc}
The source handling mechanism can accommodate three sources that are
positioned one behind the other on the beam axis, a line connecting
the source region and the center of the axis of rotation of the first
crystal.  The source material was placed in thin-walled graphite
holders that are supported on ``V''s for precise positioning.  Two
sizes of graphite holders were used.  For the two chlorine and first
sulfur measurements, the inside volume of the source holder was 2 mm
$\times$ 50 mm $\times$ 25 mm with the 2 mm $\times$ 50 mm surface
facing the spectrometer.  For the second sulfur and silicon
measurement, the inside volume was 3.5 mm $\times$ 50 mm $\times$ 25
mm with the 3.5 mm $\times$ 50 mm surface facing the spectrometer.
The neutron flux at the source position is $\approx 5.5 \times
10^{14}$~cm$^{-2}$\,s$^{-1}$.  In Table~\ref{tab:sources} the sources,
the source masses, and the estimated activities for each of the
measurements are given.  The sources that are used are extremely
active to compensate for the small effective solid angle (high
resolution) of the spectrometer.

\begin{table*}
  \caption{\label{tab:sources}Sources, source masses, and estimated
  activities for each of the measurements.}
\begin{ruledtabular}
\begin{tabular}{l....}
\multicolumn{1}{c}{Nuclide} & \multicolumn{1}{c}{Measurement} &
        \multicolumn{1}{c}{Source} & \multicolumn{1}{c}{Source mass} &
        \multicolumn{1}{c}{Estimated activity} \\
        & \multicolumn{1}{c}{dates} & & \multicolumn{1}{c}{(g)} &
        \multicolumn{1}{c}{($\times$ 10$^{13}$ Bq)} \\
        \hline
        $^{29}$Si & \multicolumn{1}{c}{Oct--Nov 2000} &
        \multicolumn{1}{c}{Si single crystal} & 13.4 & 2.6 \\
        $^{33}$S & \multicolumn{1}{c}{Sept--Oct 1998} &
        \multicolumn{1}{c}{ZnS single crystal} & 8.1 & 1.4 \\
        $^{33}$S & \multicolumn{1}{c}{April--May 1999} &
        \multicolumn{1}{c}{ZnS polycrystal} & 16.8 & 2.9 \\
        $^{36}$Cl & \multicolumn{1}{c}{Feb 1997} &
        \multicolumn{1}{c}{NaCl} & 4.5 & 85.4 \\
        $^{36}$Cl & \multicolumn{1}{c}{Sept 1997} &
        \multicolumn{1}{c}{NaCl} & 4.5 & 85.4 \\
\end{tabular}
\end{ruledtabular}
\end{table*}

\subsection{Spectrometer}
\label{subsec:spect}
The critical component of the gamma-ray facility is a double flat
crystal spectrometer that has three unique capabilities that are very
important for the accurate measurement of pico-meter wavelengths.
First, the diffraction crystals are highly perfect specimens whose
lattice spacing is measured on a scale consistent with the SI
definition of the meter.  Second, the diffraction angles are measured
with sensitive Michelson angle interferometers which are calibrated
using an optical polygon \cite{kess01}.  The angle calibration is
based on the fact that the sum of the external angles of the polygon
equals $360^{\circ}$.  Third, the gamma-ray beam collimation is
sufficient to permit the measurement of very small diffraction angles
($< 0.05^{\circ}$).  The GAMS4 facility is a precision metrology
laboratory with the usual attention to vibration isolation,
temperature control, and environmental monitoring.

Because descriptions of the profile recording and the angle
calibration that follow assume some knowledge of the angle measuring
system of the spectrometer, a few details concerning the angle
interferometers are provided.  Each of the two Michelson angle
interferometers contains two corner cube retro-reflectors that are
rigidly attached to the crystal rotation table.  As the crystal
rotates, the path length of one arm of the interferometer increases
while the path length of the other arm decreases.  The angles are
measured in whole and fractional interferometer fringes with 1 fringe
$\approx 7.8 \times 10^{-7}$\,rad.  An interferometer fringe can be
divided into $\approx 1000$ parts ($\approx 7.8 \times
10^{-10}$\,rad).  The arm that supports the retro-reflectors is made
out of low expansion invar in order to reduce the temperature
dependence of the angle interferometer.  Nevertheless, the temperature
of the corner cube arm must be taken into account when converting
interferometer fringe values into angles (see Sec.~\ref{sec:angcal}).
Because the interferometers are in the laboratory environment, all
angle fringe measurements must be corrected to standard pressure,
temperature, and humidity conditions.

\subsection{Crystals}
\label{subsec:cryst}
The gamma-rays measured in this study were diffracted by nearly
perfect silicon crystals used in transmission geometry.  All of the
crystals were cut so that the (220) family of planes was available for
diffraction and have a shape and mounting identical to the crystal
shown in Fig.~7, Ref.~\onlinecite{kess01}.  Because of the large
spread in energy of the gamma-rays, two different sets of crystals
were used.  The first set, called ILL2.5, consisted of two crystals of
nearly equal thickness ($\approx 2.5$ mm), and are the same crystals
used for the measurement of the deuteron binding energy \cite{kess99}.
The raw material for these crystals was obtained from the Solar Energy
Research Institute \cite{disclaimer}.  The second set, called
ILL4.4\&6.9, consisted of two crystals of unequal thickness ($\approx
4.41$ mm and $\approx 6.95$ mm) manufactured from raw material
obtained from Wacker \cite{disclaimer}.  The ILL2.5 crystals were used
for the lower energy lines (S 841 keV and 2380 keV and the Cl 517 keV,
786 keV, and 1951 keV lines) and initial measurements of the Cl 6111
keV line.  The ILL4.4\&6.9 crystals were used for the higher energy
lines (Si 3539 keV and 4934 keV, S 5421 keV, and Cl 6111 keV lines).
The dependence of integrated reflectivity on crystal thickness and
energy is discussed in more detail in Section 11,
Ref.~\onlinecite{kess01}.  Each line was measured in at least two
orders that were chosen on the basis of high reflectivity and small
diffraction width.  In Table~\ref{tab:orders} the crystal
configurations that were used for each energy are given along with the
nominal Bragg angles.

\begin{table*}
  \caption{\label{tab:orders}Crystals, crystal orders, and nominal
  Bragg angles for the various energies.}
\begin{ruledtabular}
\begin{tabular}{lr.....}
\multicolumn{1}{c}{Nuclide} & \multicolumn{1}{c}{Energy} &
        \multicolumn{1}{c}{Crystals} & \multicolumn{2}{c}{A crystal} &
        \multicolumn{2}{c}{B crystal} \\
        & \multicolumn{1}{c}{(keV)} & & \multicolumn{1}{c}{order-m} &
        \multicolumn{1}{c}{$\text{Bragg angle}^{\circ}$} &
        \multicolumn{1}{c}{order-n} & \multicolumn{1}{c}{$\text{Bragg
        angle}^{\circ}$} \\
      \hline
      $^{29}$Si & 3539 & \multicolumn{1}{c}{ILL4.4\&6.9} & 1 & 0.052 &
        \multicolumn{1}{c}{2,-2} & 0.105 \\
        & 3539 & \multicolumn{1}{c}{ILL4.4\&6.9} & 1 & 0.052 &
        \multicolumn{1}{c}{3,-3} & 0.157 \\
        & 4934 & \multicolumn{1}{c}{ILL4.4\&6.9} & 1 & 0.037 & 1 &
        0.037 \\
        & 4934 & \multicolumn{1}{c}{ILL4.4\&6.9} & 1 & 0.037 &
        \multicolumn{1}{c}{2,-2} & 0.075 \\
        \hline
        $^{33}$S & 841 & \multicolumn{1}{c}{ILL2.5} & 1 & 0.220 &
        \multicolumn{1}{c}{1,-1} & 0.220 \\
        & 841 & \multicolumn{1}{c}{ILL2.5} & 1 & 0.220 &
        \multicolumn{1}{c}{3,-3} & 0.660 \\
        & 2380 & \multicolumn{1}{c}{ILL2.5} & 1 & 0.078 & 1 & 0.078 \\
        & 2380 & \multicolumn{1}{c}{ILL2.5} & 1 & 0.078 &
        \multicolumn{1}{c}{2,-2} & 0.155 \\
        & 2380 & \multicolumn{1}{c}{ILL2.5} & 1 & 0.078 & $-3$ & 0.233
        \\
        & 5421 & \multicolumn{1}{c}{ILL4.4\&6.9} & 1 & 0.034 & 1 &
        0.034 \\
        & 5421 & \multicolumn{1}{c}{ILL4.4\&6.9} & 1 & 0.034 &
        \multicolumn{1}{c}{2,-2} & 0.068 \\
        & 5421 & \multicolumn{1}{c}{ILL4.4\&6.9} & 1 & 0.034 &
        \multicolumn{1}{c}{3,-3} & 0.102 \\
        \hline
        $^{36}$Cl & 517 & \multicolumn{1}{c}{ILL2.5} & 2 & 0.716 &
        \multicolumn{1}{c}{2,-2} & 0.716 \\
        & 517 & \multicolumn{1}{c}{ILL2.5} & 2 & 0.716 &
        \multicolumn{1}{c}{3,-3} & 1.074 \\
        & 786 & \multicolumn{1}{c}{ILL2.5} & 1 & 0.235 &
        \multicolumn{1}{c}{1,-1} & 0.235 \\
        & 1951 & \multicolumn{1}{c}{ILL2.5} & 1 & 0.095 &
        \multicolumn{1}{c}{2,-2} & 0.190 \\
        & 1951 & \multicolumn{1}{c}{ILL2.5} & 2 & 0.190 &
        \multicolumn{1}{c}{2,-2} & 0.190 \\
        & 6111 & \multicolumn{1}{c}{ILL2.5, ILL4.4\&6.9} & 1 & 0.030 &
        1 & 0.030 \\
        & 6111 & \multicolumn{1}{c}{ILL2.5, ILL4.4\&6.9} & 1 & 0.030 &
        2 & 0.061 \\
        & 6111 & \multicolumn{1}{c}{ILL2.5, ILL4.4\&6.9} & 1 & 0.030 &
        \multicolumn{1}{c}{3,-3} & 0.091 \\
\end{tabular}
\end{ruledtabular}
\end{table*}

In the past our approach to the determination of the unknown lattice
spacing of diffraction crystals combined two types of crystal lattice
spacing measurements: 1) absolute lattice parameter measurements in
which the lattice parameter of a particular Si crystal is compared to
the wavelength of an ${}^{127}$I$_{2}$ stabilized laser and 2) lattice
comparison (relative) measurements in which the small lattice spacing
difference between known and unknown crystal samples was measured. 
Absolute lattice parameter measurements have been published by
researchers at the Physikalisch-Technische Bundesanstalt (PTB) in 1981
\cite{beck81}, at the Istituto di Metrologia ``G. Colonnetti'' (IMGC)
in 1994 \cite{basi94}, and at the National Measurement Institute of
Japan (NMIJ) in 1997 \cite{naka97}.  Lattice comparison measurements
have been made at the PTB \cite{wind90} and the National Institute of
Standards and Technology (NIST) \cite{kess94, kess97}.  In
Ref.~\onlinecite{kess99} the above approach applied to the
determination of the lattice parameter of the ILL2.5 crystals is
described in detail using the data that was available in 1999. 
Although an early 2004 publication contains improved absolute lattice
parameter measurements from IMGC and NMIJ \cite{cava03}, the authors
have since published an erratum and advised us not to use these
results \cite{cava04}.

One of the authors of the 1998 and 2002 CODATA Recommended Values of
the Fundamental Physical Constants has recommended an alternate
approach for determining lattice parameters values for the crystals
used in these measurements \cite{private-taylor}.  In the adjustment
of the fundamental physical constants, absolute and relative lattice
parameter measurements are used in a consistent way to arrive at
recommended output values.  One of the output values is the lattice
parameter of an ideal single crystal of naturally occurring Si free of
impurities and imperfections.  Since the relative lattice parameter
measurements connecting the ILL2.5 crystals to the PTB, IMGC, and NMIJ
absolute lattice parameter crystals in Ref.~\onlinecite{kess99} are
included in the input data, the value of the lattice parameter of the
ILL2.5 crystal is an unpublished output of the adjustment process.  In
the 2002 CODATA adjustment \cite{mohr02} only the NMIJ absolute
lattice parameter value was used as input based on preliminary
measurements that were eventually published in Ref.~\onlinecite{cava03}.
Because these measurements are now known to be in error, the best
estimate for the lattice parameter of the ILL2.5 crystal must be taken
from the 1998 CODATA adjustment \cite{mohr00} and is d(220) ILL2.5 =
$1.920155822(57) \times 10^{-10}$ m at $\vartheta = 22.5~^{\circ}$C in
vacuum (relative uncertainty of $3.0 \times 10^{-8}$).  We prefer to
arbitrarily increase the relative uncertainty to $5.0 \times 10^{-8}$
to account for the present inconsistency of absolute and relative
lattice parameter results and the variation of the lattice spacing
within the raw material from which the crystals are manufactured.

To obtain a value for the lattice parameter of the ILL4.4\&6.9
crystal, the NIST lattice comparison facility was used to measure the
lattice spacing difference between the ILL2.5 and ILL4.4\&6.9
crystals.  The directly measured relative lattice parameter difference
is $(\text{ILL2.5}-\text{ILL4.4\&6.9})/\text{ILL2.5} = 4.0 \times
10^{-8}$.  By combining this relative lattice parameter measurement
with the above absolute value for d(220) ILL2.5 yields the value for
d(220) ILL4.4\&6.9 that is given in Table~\ref{tab:crystals}.  The
reasons for preferring this approach for determining lattice parameter
values over the approach that was used in Ref.~\onlinecite{kess99} are
all lattice parameter measurements included in the 1998 CODATA
adjustment are used in a consistent way to obtain a value for d(220)
ILL2.5 and only one direct lattice comparison is needed to obtain a
value for d(220) ILL4.4\&6.9.

\begin{table}
  \caption{\label{tab:crystals}Lattice spacing of ILL2.5 and
    ILL4.4\&6.9 crystals.  The ILL2.5 value is an unpublished output
    of the CODATA adjustment process.  The NIST lattice comparison
    facility was used to measure the lattice spacing difference
    between the ILL2.5 and ILL4.4\&6.9 crystals.}
  \begin{ruledtabular}
    \begin{tabular}{ll}
      $d$(220)\footnotemark[1] ILL2.5 (m) & $1.920155822(96)
      \times 10^{-10}$ \\
      \hline
      $(\text{ILL2.5}-\text{ILL4.4\&6.9})/\text{ILL2.5}$ &
      $4.0(1.0)\times10^{-8}$ \\
      \hline
      $d$(220)\footnotemark[1] ILL4.4\&6.9 (m) & $1.920155745(96)
      \times 10^{-10}$ \\
    \end{tabular}
  \end{ruledtabular}
  \footnotetext[1]{$\vartheta = 22.5~^{\circ}$C, in vacuum}
\end{table}

However, it is instructive to compare the d(220) values given in
Table~\ref{tab:crystals} to d(220) values obtained with the procedure
used in Ref.~\onlinecite{kess99}.  The relative difference between the
values for d(220) ILL2.5 given in Table~\ref{tab:crystals} and in
Ref.~\onlinecite{kess99} is $5.2 \times 10^{-8}$.  The deuteron
binding energy and the neutron mass values given in
Ref.~\onlinecite{kess99} must be corrected for this change. In order
to use the Ref.~\onlinecite{kess99} approach to determine the
ILL4.4\&6.9 lattice spacing, the ILL 4.4\&6.9 crystal was compared to
the absolute lattice parameter crystals from PTB, IMGC, and NMIJ. The
relative difference between the value for d(220) ILL4.4\&6.9 given in
Table~\ref{tab:crystals} and the value that is obtained using the
Ref.~\onlinecite{kess99} procedure is $3.9 \times10^{-8}$.  These
relative differences provide further justification for expanding the
relative uncertainty of the lattice parameter measurements to $5
\times 10^{-8}$.  In addition, the lattice parameter difference between
the ILL2.5 and ILL4.4\&6.9 crystals can be inferred from the two
d(220) values obtained with the Ref.~\onlinecite{kess99} approach.
The implied value of $(\text{ILL2.5}-\text{ILL4.4\&6.9})/\text{ILL2.5}
= 3.0 \times10^{-8}$ agrees very well with the directly measured value
given above and provides further confidence in the lattice parameter
results that are used to obtain binding energy measurements.

\subsection{Profile recording}
\label{subsec:prof}

\begin{figure}
  \includegraphics[width=3.4in]{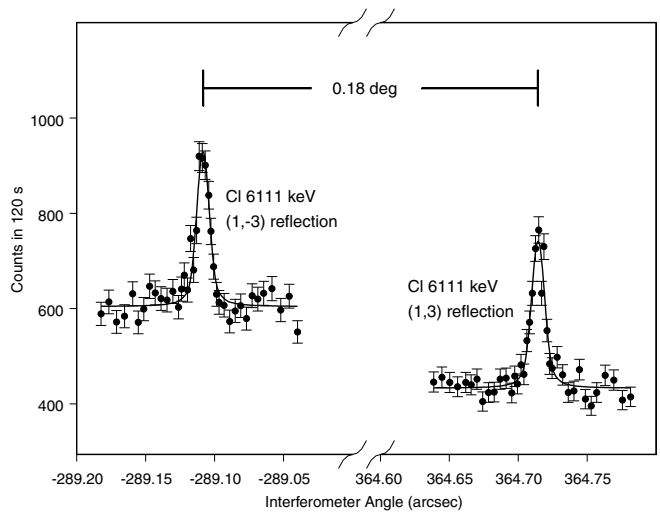}
  \caption{Two representative third order $^{36}$Cl profiles along
    with fitted curves: 6111 keV (1,-3) and (1,3) made with the thick
    crystals ILL4.4\&6.9.  The most important fit parameter is the
    profile centroid.  The difference in profile centroids is
    proportional to the Bragg angle while the numeric mean of the two
    centroids gives the offset angle between the second crystal
    diffracting planes and the angle interferometer.}
  \label{fig:profile}
\end{figure}

Gamma-ray profiles (intensity vs interferometer fringes) were recorded
for the crystal configurations given in Table~\ref{tab:orders} by
scanning the angular position of the second crystal.  The profiles
consisted of 30 to 45 points with counting times from 30 to 180
seconds per point and were scanned in both the cw and ccw directions.
For each data point the interferometer fringe value was reduced to
standard atmospheric conditions (pressure, temperature, and humidity)
\cite{birc93, birc94}.  The gamma-ray counts were accumulated in a Ge
detector-multichannel analyzer system.  The recorded profiles were
least squares fit to theoretical dynamical diffraction profiles
broadened with a Gaussian function to account for crystal
imperfections, vibrations, and thermal and recoil induced motions of
the atoms in the source.  The use of a single Gaussian function to
account for deviations of the recorded profiles from the theoretical
dynamical diffraction profiles has been shown to provide reliable peak
positions, but is not sufficient to obtain nuclear level lifetimes and
recoil velocities from profile width measurements \cite{born93}.  The
adjustable parameters in the fit are the position, intensity,
background, and Gaussian width contribution.  In the fit, the number
of counts at fringe value $i$, $n_i$, is weighted by
$(1/\sqrt{n_i})^{2}$.  The detector and fitting procedure are
described in more detail in Ref.~\onlinecite{kess01}.  Two
representative profiles are shown in Fig.~\ref{fig:profile}.  The
profile positions (in interferometer fringes), the most important
parameter for the determination of energies, are converted into
diffraction angles by using the second-axis angle calibration that is
described in the next section.

\section{Angle Interferometer Calibration}
\label{sec:angcal}

A complete description of a calibration run can be found in
Ref.~\onlinecite{kess01}.  As discussed there the formula connecting
optical fringes $f$, which are recorded with the rocking curves, and
the true interferometer angle $\theta$ is
\begin{equation}
  \label{eq:fofth}
  f = K \sin \theta + f_0\,,
\end{equation}
where $K$ is the instrument calibration constant and $f_0$ is an
electronic offset.  Both terms must be determined experimentally
through a calibration procedure.  Ideally $K$ would be invariable,
however experience shows that $K$ depends on temperature, time, 
humidity and interferometer laser and alignment.

\subsection{The Global Calibration Procedure}

\begin{figure}
  \includegraphics[width=3.4in]{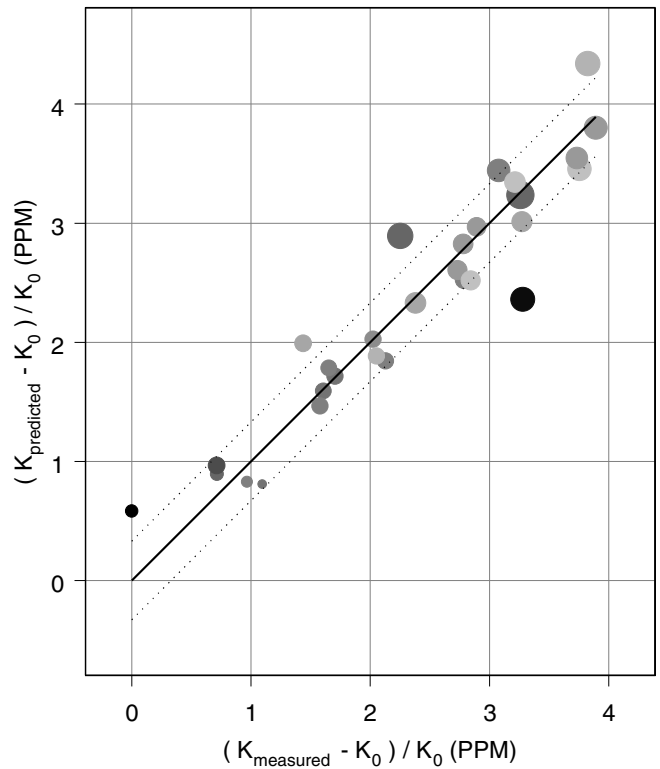}
  \caption{\label{fig:schematic}Global fit to 29 GAMS4 calibrations.
    Predictions from a five parameter global fit
    [Eq.~(\ref{eq:calib})] are plotted versus measured calibration
    values.  The standard deviation of the relative residuals given in
    the last column of Table~\ref{tab:calibrations} is $0.33
    \times 10^{-6}$.  The dotted lines correspond to $\pm 0.33 \times
    10^{-6}$ about the fit.  Across the data sets the temperature
    ranges from 25.5 to $26.8~^{\circ}$C corresponding to a $1.9
    \times 10^{-6}$ relative change in K, the relative humidity ranges
    from 0.21 to 0.58 corresponding to a $3.0 \times 10^{-6}$ relative
    change in K, and 4.6 years separates the first and last
    measurements corresponding to a $2.0 \times 10^{-6}$ relative
    change in K.  The size of the plotted points is proportional to
    the humidity while darker (lighter) points represent lower
    (higher) temperatures.}
\end{figure}

Between September 1997 and May 2002, a period spanning the
measurements described in this paper, the GAMS4 spectrometer was
calibrated 29 times.  Table~\ref{tab:calibrations} lists each of these
calibrations.  The experimentally determined calibration constants
(column 5) fit well to a linearized equation
\begin{eqnarray}
  \label{eq:calib}
  K = &K_0 + K_\vartheta ( \vartheta - 26 ) + K_d ( d - 800 )  + K_h (
 h - 0.35 ) \nonumber\\
 &  + K_{\text{laser}} \times \begin{cases}
    -1 & \text{if $d<700$},\\
    1 & \text{if $d>700$}\end{cases}
\end{eqnarray}
where $K = K(\vartheta,d,h)$ is the desired calibration constant,
$\vartheta$ is the corner cube arm temperature ($^{\circ}$C), $h$ is
the relative humidity, $d$ is the number of days after 8/31/1997, and
$d=700$ corresponds to 8/1/1999 when the interferometer laser was
replaced.  A least squares fit to the data in
Table~\ref{tab:calibrations} gives
\begin{align}\label{eq:coef}
  K_0 & = 5133462.12 \pm 0.59\notag\\
  K_\vartheta & = 7.5 \pm 1.3\; /{}^{\circ}\text{C}\notag\\
  K_d & = -0.0059 \pm 0.0013\notag\\
  K_h & = 41.6 \pm 3.9\notag\\
  K_{\text{laser}} & = 3.13 \pm 0.57
\end{align}
where $\vartheta_{\text{avg}}=26~^{\circ}$C, $h_{\text{avg}}=0.35$,
and $d_{\text{avg}}=800$ (11/9/1999) are average conditions around
which the fit is made.  This fit is plotted in
Fig.~\ref{fig:schematic}.  The relative standard deviation of the
residuals (column 7) is $0.33 \times 10^{-6}$.  Eqs.~(\ref{eq:fofth}),
(\ref{eq:calib}), and (\ref{eq:coef}) along with values for the
profile centroid in fringes, the mean corner cube arm temperature, the
mean humidity, and the mean wall time can be used to extract a profile
centroid in radians from a single data file.  This procedure uses all
available calibration data to determine one set of coefficients
($K_{0}, K_{\vartheta}, K_{d}, K_{h}, K_{\text{laser}}$) that are
assumed valid for all of the data presented in this paper.  In the
remainder of this paper this procedure is referred to as the Global
Calibration.

\begin{table*}
  \caption{\label{tab:calibrations}GAMS4 angle calibrations.  Each
    line represents a unique calibration of the spectrometer.  Column
    1 gives the date of the calibration, column 2 the nuclide being
    measured, column 3 the calibration-average corner cube arm
    temperature ($^{\circ}$C), column 4 the calibration-average
    relative humidity, column 5 the measured instrument calibration
    constant $K$, column 6 the measured calibration zero, and column 7
    the residuals from a five parameter fit to $K$.  The horizontal
    line indicates the time when the interferometer laser was replaced
    causing a $(1.22 \pm 0.22) \times 10^{-6}$ one time fractional
    shift in K.  ``CC'' denotes corner cube.}
  \begin{ruledtabular}
    \begin{tabular}{lc.....}
      & Nuclide & \multicolumn{1}{c}{Avg CC arm} &
      \multicolumn{1}{c}{Avg relative} &&& \multicolumn{1}{c}{Fit}\\
      Date & measured & \multicolumn{1}{c}{temp
      ($^{\circ}$C)} & \multicolumn{1}{c}{humidity} &
      \multicolumn{1}{c}{$K$} & \multicolumn{1}{c}{$f_0$} &
      \multicolumn{1}{c}{rel dev$\times 10^6$} \\ 
      \hline
      09/27/1997 & $^{36}$Cl & 25.595 & 0.495 & 5133471.3 & 588.3 &
      -0.92 \\
      03/08/1998 & & 26.272 & 0.317 & 5133463.2 & -747.7 & 0.01 \\
      03/26/1998 & & 26.266 & 0.209 & 5133460.1 & -721.1 & -0.28 \\
      03/30/1998 & & 26.256 & 0.308 & 5133462.7 & -721.2 & -0.01 \\
      09/26/1998 & $^{33}$S & 26.716 & 0.481 & 5133473.7 & -547.5 &
      -0.30 \\
      10/21/1998 & $^{33}$S & 26.607 & 0.323 & 5133461.8 & -547.7 &
      0.55 \\
      11/05/1998 & $^{33}$S & 26.308 & 0.314 & 5133462.6 & -203.1 &
      -0.11 \\
      11/17/1998 & $^{33}$S & 26.199 & 0.265 & 5133458.1 & -202.9 &
      0.18 \\
      04/13/1999 & $^{33}$S & 26.041 & 0.323 & 5133458.1 & -186.9 &
      0.25 \\
      04/30/1999 & $^{33}$S & 26.194 & 0.536 & 5133466.0 & -186.9 &
      0.64 \\
      05/07/1999 & $^{33}$S & 26.173 & 0.583 & 5133471.2 & -68.2 &
      -0.02 \\
      \hline
      09/17/1999 & & 26.492 & 0.464 & 5133474.4 & -268.2 & -0.09 \\
      09/19/1999 & & 26.486 & 0.434 & 5133473.6 & -268.1 & -0.18 \\
      09/25/1999 & & 26.648 & 0.503 & 5133474.1 & -255.5 & 0.52 \\
      10/05/1999 & & 26.354 & 0.334 & 5133468.8 & -254.5 & -0.26 \\
      10/15/1999 & & 26.348 & 0.449 & 5133470.3 & -249.5 & 0.37 \\
      10/18/1999 & & 26.477 & 0.369 & 5133469.3 & -249.4 & 0.08 \\
      12/03/1999 & & 25.493 & 0.257 & 5133454.5 & -249.5 & 0.59 \\
      07/06/2000 & & 26.545 & 0.399 & 5133471.3 & -183.5 & -0.26 \\
      07/18/2000 & & 26.474 & 0.390 & 5133468.7 & -183.4 & 0.05 \\
      10/15/2000 & $^{29}$Si & 26.779 & 0.412 & 5133471.0 & -253.2 &
      0.13 \\
      10/22/2000 & $^{29}$Si & 26.490 & 0.374 & 5133468.5 & -192.4 &
      -0.12 \\
      11/05/2000 & $^{29}$Si & 26.313 & 0.314 & 5133465.4 & -240.8 &
      -0.28 \\
      11/14/2000 & $^{29}$Si & 26.409 & 0.320 & 5133464.8 & -262.2 &
      0.00 \\
      11/28/2000 & $^{29}$Si & 26.302 & 0.312 & 5133462.9 & -276.1 &
      0.14 \\
      10/28/2001 & & 26.725 & 0.374 & 5133469.1 & -211.8 & -0.32 \\
      11/06/2001 & & 26.619 & 0.316 & 5133465.0 & -250.0 & -0.16 \\
      11/12/2001 & & 26.347 & 0.236 & 5133459.4 & -250.0 & -0.13 \\
      05/19/2002 & & 26.551 & 0.411 & 5133466.7 & -263.4 & -0.05 \\
    \end{tabular}
  \end{ruledtabular}
\end{table*}

The dependence of $K$ on temperature and time is not unexpected as
the physical dimensions of the interferometer corner cube arm can vary
with temperature and time in roughly the amounts seen
\cite{berg97,bert76}.  The data reveals no dependence of $K$ on
atmospheric pressure which is as expected since the interferometer
fringe values are reduced to standard atmospheric conditions and the
small pressure changes are not likely to alter the dimensions of the
interferometer.  The relative magnitude of the dependence on
interferometer laser ($\approx 1.22 \times 10^{-6}$) is somewhat
larger than the expected variability in the laser wavelength and is
likely due to interferometer alignment.  The dependence of $K$ on
humidity is unexpected and no definitive cause has been established. 
In an effort to better understand the dependence of the calibration on
the environment, more frequent calibrations were performed as the
measurements progressed.  As is evident from Fig.~\ref{fig:schematic},
there are significant differences between some values of
$K_{\text{measured}}$ and $K_{\text{fit}}$.  These differences
prompted us to consider alternate calibration procedures.

\subsection{Other Calibration Procedures}
\label{sec:othercalib}

For the later measurement cycles, multiple calibrations were
performed: $^{33}$S 1998 - 4, $^{33}$S 1999 - 3 and $^{28}$Si 2000 -
5.  Since these calibrations were dispersed throughout the measurement
cycle, it is possible to interpolate between the measured calibrations
using spline fitting to obtain calibration constant values for
converting profile data in fringes to angles.  This procedure assumes
that each of the recorded angle calibrations is valid and the angle
calibration varies smoothly with time during the measurement cycle.
Although the individual calibration values within a measurement cycle
are dependent on temperature, humidity, and day number, the spline
fitting uses only the individual calibration values and assumes no
explicit dependence on temperature, humidity or day number.  We have
called this calibration procedure the Spline Calibration.  It uses
only the calibration information obtained during a given measurement
cycle to analyze the wavelength data recorded in that cycle.  The
$^{33}$S and $^{29}$Si data were analyzed using the spline calibration
and the relative difference between wavelengths obtained with the
global and spline calibration procedures is $\approx 0.2 \times
10^{-6}$.  Since only one calibration was recorded during the
$^{36}$Cl 1997 measurement cycle, the spline calibration could not be
applied to this data.  This led us to a third calibration procedure
called the Local-Global Calibration.

In the local-global calibration, the calibrations that were performed
in a particular measurement cycle are given a more significant role in
the determination of $K(\vartheta,d,h)$ for that measurement cycle.
First, $K_{\vartheta}$, $K_{d}$, $K_{h}$, and $K_{\text{laser}}$ are
taken as given in Eq.~(\ref{eq:coef}) since these coefficients are not
expected to change with time and a global fit provides the best
estimate of these coefficients.  Next the subset of calibrations
performed in each measurement cycle are fit to the equation
\begin{eqnarray}
  \label{eq:MethodB}
  K = &K_0 + 7.5 ( \vartheta - 26 ) -0.0059 ( d - 800 )
 \nonumber\\
 &   + 41.6 ( h - 0.35 ) + 3.13 \times \begin{cases}
    -1 & \text{if $d<700$},\\
    1 & \text{if $d>700$}\end{cases}
\end{eqnarray}
to obtain independent values of $K_{0}$ for each measurement cycle.
All of the $^{33}$S and $^{29}$Si data were analyzed using the
local-global calibration procedure and the relative difference between
wavelengths obtained with global and local-global calibration
procedures is $\approx 0.3 \times 10^{-6}$.  The $^{36}$Cl data were
also analyzed using the local-global calibration procedure and show
significantly larger relative differences ($\approx 0.9 \times
10^{-6}$) between wavelengths obtained with the global and
local-global calibration procedures.  These large differences result
from recording only one calibration during the $^{36}$Cl measurement
cycle and from the fact that this particular calibration is the most
discrepant point in the global fit of Fig.~\ref{fig:schematic}.

Consideration of these alternate calibration procedures did not
provide sufficient evidence to make a ``best'' calibration procedure
choice. All of the wavelength and energy values reported in this paper
were obtained using the global calibration procedure. Our reasons for
choosing this procedure are 1) all of the available data can be
analyzed using one procedure and 2) this procedure makes the maximum
use of the available calibration data.

\subsection{Calibration Uncertainty}
\label{sec:calibuncer}

Although consideration of three calibration procedures did not lead to
a clear ``best'' procedure, this exercise does provide an estimate of
the calibration uncertainty.  The variation of wavelengths obtained
with the global, spline and local-global calibration procedures
suggests a relative calibration uncertainty of $\text{0.2--0.3} \times
10^{-6}$.  A second estimate of the calibration uncertainty is
available from the fit used in the global calibration procedure.  The
standard deviation of the relative residuals given in
Table~\ref{tab:calibrations} and shown in Fig.~\ref{fig:schematic} is
$0.33 \times 10^{-6}$ and provides a measure of the quality of the fit
and the uncertainty of this calibration procedure.  A third estimate
of the calibration uncertainty is the variation of the wavelength
values obtained in different measurement cycles.  In this approach
lower energy (larger angles) intense transitions must be used because
higher energy (smaller angles) weak transitions have a statistical
uncertainty that masks the calibration uncertainty.  The 841 keV line
in $^{33}$S was measured in 1998 and 1999 and shows a relative excess
variation of $0.38 \times 10^{-6}$ if the global calibration procedure
is used and $0.16 \times 10^{-6}$ if the spline calibration is used.
In addition, this instrumental set up was used to measure an intense
line that does not contribute to the binding energy determinations
presented here, namely the 816 keV line in $^{168}$Er.  This line was
measured in two different measurement cycles, Oct-Nov 2000 and Nov
2003 and shows a relative excess variation of $0.45 \times 10^{-6}$
and $0.3 \times 10^{-6}$ for the global and spline calibrations,
respectively.  Although it is difficult to obtain a rigorous
calibration uncertainty from these three estimates, we choose to use
the above values to arrive at a slightly conservative relative
calibration uncertainty of $0.4 \times 10^{-6}$.  This calibration
uncertainty will be combined with the statistical and other systematic
uncertainties to obtain final wavelength uncertainties.

\section{Wavelength measurements}
\label{sec:bragg}
Wavelengths are determined by combining a sequence of profile angle
measurements.  First the profile centroids and uncertainties in
interferometer fringes are converted into angles via
Eq.~(\ref{eq:fofth}).  The calibration constant $K$ that appears in
this equation has been appropriately corrected for the corner cube arm
temperature, the relative humidity, and time.  Each of these angles is
next corrected for vertical divergence as discussed in
Ref.~\onlinecite{schn65}.  Typically these corrections are between 2
and $4 \times 10^{-7}$ in relative size.  For a given energy and
configuration the profiles are recorded with the first crystal in a
fixed position and the second crystal sequentially in a more and less
dispersive position (see Fig.~\ref{fig:profile}).  In addition, the
data recording sequence includes both cw and ccw rotation of the
second crystal.  To more concretely illustrate how wavelengths are
determined, we use the $^{33}$S 841 keV (1,$-3$), (1,3) measurement as
an example.  A group of four profiles recorded in the sequence (1,$-3$
cw), (1,3 cw), (1,3 ccw), (1,$-3$ ccw) is used to determine a
wavelength value.  The four angles associated with the four profiles
are fit with the equation
\begin{equation}
  \label{eq:bragg}
  \theta(n,t,\vartheta) =
  \arcsin\left(\frac{n\lambda_{\text{meas}}}{2d(\vartheta)}\right) +
  \theta_{0}(t)\,,
\end{equation}
where $n$ is the diffraction order, $t$ is the time, $\vartheta$ is
the crystal temperature, $\lambda_{\text{meas}}$ is the sought after
wavelength, $d(\vartheta)$ is the lattice spacing at crystal
temperature $\vartheta$ and $\theta_{0}(t)$ represents the potentially
time dependent angular offset between the second crystal diffracting
planes and the angle interferometer.  The symbol
$\lambda_{\text{meas}}$ is introduced to indicate that wavelengths
determined using this equation are laboratory measured wavelengths
(not corrected for recoil); $d(\vartheta)$ is given by
\begin{equation}
  \label{eq:doft}
  d(\vartheta) = d_{\text{$22.5~^{\circ}$C, atm}} ( 1 + 2.56 \times
  10^{-6} ( \vartheta - 22.5 ) )\,,
\end{equation}
where $d_{\text{$22.5~^{\circ}$C, atm}}$ is the lattice spacing at
$22.5~^{\circ}$C and atmospheric pressure.  The lattice parameter
measurements given in Table~\ref{tab:crystals} are specified for
vacuum.  To obtain the value at the pressure present in the reactor
hall ($p \approx 0.987$ atmospheres) it is necessary to use the
following transformation
\begin{equation}
  \label{eq:vacair}
  d_{\text{$22.5~^{\circ}$C}}(p) = d_{\text{$22.5~^{\circ}$C, vac}} (
  1 - \epsilon p )\,,
\end{equation}
where p is the pressure.  In this equation $\epsilon=0.3452 \times
10^{-6}$/atmosphere \cite{nye57,mcsk53}.

In the fit to Eq.~(\ref{eq:bragg}), the three parameters are the
wavelength, a constant angular offset, and a linear temporal offset
term (i.e\@., $\theta_{0}(t) = a + b t$).  Each of the four angles
$\theta_{i}$ is weighted by $(1/\sigma_{\theta_{i}})^2$ where
$\sigma_{\theta_{i}}$ is the profile centroid uncertainty.  This
sequence of profiles is repeated multiple times in at least two
different sets of orders (here 29 instances of (1,$-3$), (1,3); 21
instances of (1,$-1$), (1,1); and 1 instance of (1,$-3$), (1,$-1$),
(1,1), (1,3)).  An uncertainty equal to the standard deviation of the
wavelength determinations in a unique set of orders is assigned to
each wavelength determination derived from that set of orders.  The
average wavelength is the weighted mean of all the individual
determinations.  We find no evidence for an order dependent effect in
the wavelength data.

Table~\ref{tab:braggang} gives the nuclide, the nominal energy, the
crystals, the total number of Bragg angle measurements and mean
wavelength along with the statistical uncertainties in parentheses.
The sulfur transitions appear twice because they were measured in 1998
and 1999.

\begin{table*}
  \caption{\label{tab:braggang} Values of the measured
  wavelengths for the transitions included in the binding energy
  determinations.  The uncertainties are statistical only.}
\begin{ruledtabular}
\begin{tabular}{lrcclc}
\multicolumn{1}{c}{Nuclide} & \multicolumn{1}{c}{Energy} &
Crystals & Number of &
\multicolumn{1}{c}{$\lambda_{\text{meas}} \times 10^{12}$} & relative \\
& \multicolumn{1}{c}{(keV)} & & Bragg angle &
\multicolumn{1}{c}{(m)} & uncertainty \\
& & & measurements & & $\times 10^6$ \\
\hline
$^{29}$Si & 3539 & ILL4.4\&6.9 & 42 &
0.350340126(44) & 0.13 \\
& 4934 & ILL4.4\&6.9 & 147 & 0.25128808(18) & 0.70 \\
\hline
$^{33}$S & 841\rlap{\footnotemark[1]} & ILL2.5 & 26 & 1.47429306(14) &
0.09 \\
& 2380\rlap{\footnotemark[1]} & ILL2.5 & 18 & 0.52103852(30) & 0.58 \\
& 5421\rlap{\footnotemark[1]} & ILL4.4\&6.9 & 42 & 0.228730970(63) &
0.27 \\
& 841\rlap{\footnotemark[2]} & ILL2.5 & 25 & 1.47429225(12) & 0.08 \\
& 2380\rlap{\footnotemark[2]} & ILL2.5 & 31 & 0.52103905(23) & 0.43 \\
& 5421\rlap{\footnotemark[2]} & ILL4.4\&6.9 & 30 & 0.22873089(13) &
0.56 \\
\hline
$^{36}$Cl & 517 & ILL2.5 & 15 & 2.39782393(10) & 0.04 \\
& 786 & ILL2.5 & 2 & 1.57681233(83) & 0.53 \\
& 1951 & ILL2.5 & 12 & 0.63544928(10) & 0.16 \\
& 6111 & ILL2.5, ILL4.4\&6.9 & 17 & 0.20288757(10) & 0.50 \\
\end{tabular}
\end{ruledtabular}
\footnotetext[1]{Sept--Oct 1998}
\footnotetext[2]{April--May 1999}
\end{table*}

In Table~\ref{tab:wavelen}, final measured wavelength values and
uncertainties are reported (column 3).  Four additional sources of
uncertainty have been added in quadrature to the statistical
uncertainties given in Table~\ref{tab:braggang}.  These are the
calibration uncertainty discussed in Sec.~\ref{sec:angcal} and
uncertainties associated with the crystal temperature, the vertical
divergence of the gamma-ray beam, and the measured lattice spacing.  A
relative calibration uncertainty of $0.4 \times 10^{-6}$ is applied to
each energy listed in Table~\ref{tab:braggang}.  To obtain final
sulfur wavelengths and uncertainties it is necessary to add the
calibration and statistical uncertainties in quadrature for each
sulfur transition before statistically combining the 1998 and 1999
values.  This has the effect of reducing the calibration uncertainty
for the sulfur lines because they were measured twice.  In the case of
binding energies where one sums transition energies, the calibration
uncertainty is applied to the sum rather than to the constituent
transitions; and again, two binding energies are combined to arrive at
a final sulfur binding energy.  The other three uncertainties are
applied to the final transition energies or to the binding energies.
An uncertainty in the crystal temperature measurement of
0.05~$^{\circ}$C contributes a relative uncertainty of $0.1 \times
10^{-6}$.  The vertical divergence uncertainty accounts for the
possible misalignment of the gamma-ray beam with respect to the plane
of dispersion of the spectrometer.  A misalignment of 5 mm over a
distance of 15 m leads to a relative wavelength uncertainty of $0.06
\times 10^{-6}$.  This uncertainty is approximately 20\% of the size
of the correction.  From Sec.~\ref{subsec:cryst} and
Table~\ref{tab:crystals}, the relative crystal lattice spacing
uncertainty is $0.05 \times 10^{-6}$.

\section{Binding energy measurements}
\label{sec:bindenergy}
In column 4 of Table~\ref{tab:wavelen} the corresponding energy
equivalent values $E_{\text{meas}}$ are given, where the conversion
factor $1\,\text{m}^{-1} = 1.23984191(11) \times 10^{-6}$ eV was used
\cite{mohr02}.

\begin{table*}
  \caption{\label{tab:wavelen}Measured and recoil corrected
      (transition) wavelengths and energies.  Measured values from
      Ref.~\onlinecite{kess99} for ${}^{2}$H, which have been adjusted
      for the adjusted value of d(220) ILL2.5, are included for
      convenience.}
\begin{ruledtabular}
\begin{tabular}{lrl.l.c}
  \multicolumn{1}{c}{Nuclide} & \multicolumn{1}{c}{Energy}
  & \multicolumn{1}{c}{$\lambda_{\text{meas}} \times 10^{12}$}
  & \multicolumn{1}{c}{$E_{\text{meas}}$} &
  \multicolumn{1}{c}{$\lambda_{\text{trans}} \times 10^{12}$} &
  \multicolumn{1}{c}{$E_{\text{trans}}$} & wavelength relative \\
  & \multicolumn{1}{c}{(keV)} &
  \multicolumn{1}{c}{(m)} & \multicolumn{1}{c}{(eV)}
  & \multicolumn{1}{c}{(m)} &
  \multicolumn{1}{c}{(eV)} & uncertainty \\
  & & & & & & $\times 10^6$ \\
  \hline
  ${}^{29}$Si & 3539 & 0.35034013(15) & 3538966.3(1.6) &
  0.35031716(15) & 3539198.3(1.6) & 0.44 \\
  & 4934 & 0.25128808(21) & 4933946.3(4.0) & 0.25126511(20) &
  4934397.4(4.0) & 0.82 \\
  \hline
  ${}^{33}$S & 841 & 1.47429265(47) & 840974.08(28) & 1.47427246(47) &
  840985.60(28) & 0.32 \\
  & 2380 & 0.52103883(24) & 2379557.6(1.1) & 0.52101864(24) &
  2379649.8(1.1) & 0.47 \\
  & 5421 & 0.228730944(95) & 5420525.5(2.3) & 0.228710758(95) &
  5421003.9(2.3) & 0.42 \\
  \hline
  ${}^{36}$Cl & 517 & 2.3978239(10) & 517069.62(22) & 2.3978054(10) &
  517073.61(22) & 0.42 \\
  & 786 & 1.5768123(11) & 786296.43(53) & 1.5767938(11) &
  786305.66(53) & 0.67 \\
  & 1951 & 0.63544928(29) & 1951126.47(89) & 0.63543077(29) &
  1951183.30(89) & 0.45 \\
  & 6111 & 0.20288757(13) & 6110980.2(4.0) & 0.20286906(13) &
  6111537.6(4.0) & 0.66 \\
  \hline
  ${}^{2}$H & 2223 & 0.557671328(99) & 2223248.44(44) &
  0.557341007(99) & 2224566.10(44) & 0.18 \\
\end{tabular}
\end{ruledtabular}
\end{table*}

The measured wavelength values must be corrected for recoil to obtain
wavelength values $\lambda_{\text{trans}}$ whose corresponding
energies can be summed to obtain binding energies $S_{n}$.  To
excellent approximation we account for the recoil by the
transformation
\begin{equation}
  \label{eq:recoil}
  \frac{hc}{\lambda_{\text{trans}}} =
  \frac{hc}{\lambda_{\text{meas}}} +
  \frac{1}{2Mc^{2}}\left(\frac{hc}{\lambda_{\text{meas}}}\right)^{2}\,,
\end{equation}
where $M$ is the mass of the decaying nucleus, $h$ is the Planck
constant, and $c$ is the speed of light.  The second term in
Eq.~(\ref{eq:recoil}) accounts for the loss of energy imparted to the
recoiling nucleus.  If the decay occurs in flight, there will be a
first order Doppler effect.  It causes no shift in the central value
as long as the motion is isotropic as is expected to be the case here.
As discussed in Ref.~\onlinecite{yong99}, two additional terms appear
in Eq.~(\ref{eq:recoil}) in the case of decays from the capture state.
First, the kinetic energy of the incident neutron ($\approx
0.057$\,eV) must be subtracted from the measured gamma-ray energy.
For the three capture gamma-rays measured here (3539 keV in $^{29}$Si,
5421 keV in $^{33}$S, and 6111 keV in $^{36}$Cl), this effect is less
than $0.02 \times 10^{-6}$.  Second, there is a Doppler term if the
incident neutron comes from a particular direction.  The relative
peak-to-peak amplitude of this term is $\approx 0.6 \times 10^{-6}$
for the nuclei being discussed here.  The term vanishes if the
incident neutron direction is isotropic as is expected to be the case
here.  The relative uncertainty of the recoil correction is estimated
to be no greater than $0.01 \times 10^{-6}$.  As such it is negligible
given the current accuracy of $\lambda_{\text{meas}}$.  Values for
$\lambda_{\text{trans}}$ and the corresponding energy equivalent
$E_{\text{trans}}$ are given in Table~\ref{tab:wavelen}, column 5 and
6.

To obtain a wavelength equivalent to the binding energy,
$\lambda_{\text{be}}$, we sum the reciprocals of the
$\lambda_{\text{trans}}$ values for the transitions comprising the
binding energy cascade and then convert $\lambda_{\text{be}}$ into
atomic mass units using the conversion factor: $1\,\text{m}^{-1} =
1.3310250506(89) \times 10^{-15}$ u \cite{mohr02}.  Likewise, the
binding energy can be expressed in eV by using the $\text{m}^{-1}$ to
eV conversion factor given above.  Table~\ref{tab:bindeng} contains
values for the three binding energies of $^{29}$Si, $^{33}$S, and
$^{36}$Cl in meters, atomic mass units and electron volts.

\begin{table*}
  \caption{\label{tab:bindeng}Measured binding energies in meters (m),
    atomic mass units (u) and electron volts (eV).
        $\lambda_{\text{be}} = \frac{1}{\sum_{i}
        \frac{1}{\lambda_{\text{trans}_{i}}}}$ is the wavelength of a
        photon whose energy is equal to  the binding energy.
    Measured values from Ref.~\onlinecite{kess99} for ${}^{2}$H, which
      have been adjusted for the adjusted value of d(220) ILL2.5, are
      included for convenience.}
\begin{ruledtabular}
\begin{tabular}{l....}
  Nuclide & \multicolumn{1}{c}{$\lambda_{\text{be}} \times 10^{12}$} &
  \multicolumn{1}{c}{$S_{n} \times 10^{3}$} &
  \multicolumn{1}{c}{$S_{n}$} & \multicolumn{1}{c}{wavelength relative} \\
  & \multicolumn{1}{c}{(m)} & \multicolumn{1}{c}{(u)} &
  \multicolumn{1}{c}{(eV)} & \multicolumn{1}{c}{uncertainty} \\
  &&&& \multicolumn{1}{c}{$\times 10^{6}$} \\
  \hline
  $^{29}$Si & 0.146318275(86) & 9.0967793(53) & 8473595.7(5.0) & 0.59 \\
  $^{33}$S & 0.143472991(54) & 9.2771820(35) & 8641639.8(3.3) & 0.38 \\
  $^{36}$Cl & 0.144507180(80) & 9.2107883(51) & 8579794.5(4.8) & 0.55 \\
  ${}^{2}$H & 0.557341007(98) & 2.38816996(42) & 2224566.10(44) & 0.18 \\
\end{tabular}
\end{ruledtabular}
\end{table*}

\section{Discussion}
\label{sec:disc}

In this section we discuss the consistency of gamma-ray based and
atomic mass based binding energy measurements and compare the
gamma-ray measurements reported here with other high precision
gamma-ray measurements.  As discussed in Sec.~\ref{sec:intro},
precision atomic mass measurements can be used to determine binding
energies.  By using atomic mass values from the Atomic Mass Data
Center \cite{amdc,audi95} and the fundamental constants \cite{mohr02}
along with Eq.~(\ref{eq:ambe}), binding energies primarily based on
atomic mass measurements can be determined.  However, since the
determination of the neutron mass includes a gamma-ray measurement,
binding energy determinations based on Eq.~(\ref{eq:ambe}) are a
combination of atomic mass measurements and gamma-ray measurements.
This difficulty can be circumvented by expressing $m(n)$ in terms of
$m(\text{H})$, $m({}^{2}\text{H})$ and $S_{n}({}^{2}\text{H})$.  By
making this substitution and rearranging terms, Eq.~\ref{eq:ambe}
becomes
\begin{eqnarray}
  \label{eq:ambe1}
  m({}^{A}\text{X}) - m({}^{A+1}\text{X}) + m({}^{2}\text{H}) -
  m(\text{H}) = \nonumber \\
  S_{n}({}^{A+1}\text{X}) - S_{n}({}^{2}\text{H})\,.
\end{eqnarray}
Since the left and right sides of this equation involve only atomic
mass and gamma-ray measurements respectively, this equation is a valid
test of the consistency of high-precision atomic mass and gamma-ray
measurements.

Recently, new highly accurate values for the left hand side of this
equation have been reported for ${}^{A+1}\text{X}$ equal to ${}^{29}$Si and
${}^{33}$S \cite{private-pritchard}.  In these measurements the
cyclotron frequencies of two different ions simultaneously confined in
a Penning trap were directly compared.  The measured quantities are
the mass ratios $m[{}^{33}\text{S}^{+}]/m[{}^{32}\text{SH}^{+}]$ and
$m[{}^{29}\text{Si}^{+}]/m[{}^{28}\text{SiH}^{+}]$ from which, along
with the quantity $m({}^{2}\text{H}) - 2 m(\text{H})$, the mass
differences on the left hand side of Eq.~\ref{eq:ambe1} are derived.
In Table~\ref{tab:consistency}, column 2 the left hand side mass
differences for Si and S are given.  The relative uncertainties for
the Si and S values in column 2 are $7.0 \times 10^{-8}$ and $7.3
\times 10^{-8}$ respectively.

Values for the right hand side of this equation for ${}^{A+1}\text{X}$ equal
to ${}^{29}$Si and ${}^{33}$S follow directly from the binding
energies given in Table~\ref{tab:bindeng} and are given in
Table~\ref{tab:consistency}, column 3.  The relative uncertainties for
the Si and S values in column 3 are $8.0 \times 10^{-7}$ and $5.1
\times 10^{-7}$ respectively.

\begin{figure}
  \includegraphics[width=3.4in]{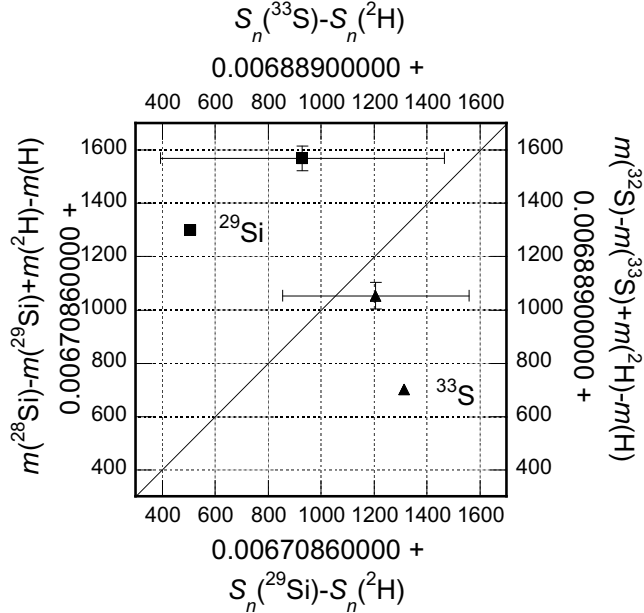}
  \caption{\label{fig:comparison}Mass differences determined in a
    Penning trap versus those determined from wavelength
    determinations.  The left (right) and bottom (top) axes correspond
    to Si (S).  The scales of the axes have been chosen so that a
    diagonal line through the plot represents exact consistency
    between Penning trap and gamma-ray measurements.  The plot shows
    that the quality of the consistency test is limited by the
    uncertainty of the gamma-ray measurements.}
\end{figure}

In Fig.~\ref{fig:comparison} the values given in
Table~\ref{tab:consistency} are plotted to show the consistency of the
atomic mass and gamma-ray measurements.  The left and bottom axes are
used for $^{29}$Si, while the top and right axes are used for
$^{33}$S. The scales of the axes have been chosen so that a diagonal
line through the plot represents exact consistency between atomic mass
and gamma-ray measurements.  The plot shows that the quality of the
consistency test is limited by the uncertainty of the gamma-ray
measurements.  For $^{29}$Si the two measurements are slightly
inconsistent (1.2~$\sigma$), while for $^{33}$S the two measurements
agree within the uncertainty (0.4~$\sigma$).

In Table~\ref{tab:consistency}, column 4 the fractional difference
between the atomic mass and gamma-ray measurements is given along with
the weighted average of the Si and S fractional differences.  These
measurements confirm the consistency of atomic mass and gamma-ray
measurements with a relative uncertainty of $4.3 \times 10^{-7}$.

\begin{table*}
  \caption{\label{tab:consistency}Consistency of high precision atomic 
  mass and gamma-ray measurements.}
\begin{ruledtabular}                                                    
\begin{tabular}{lcc.}
\multicolumn{1}{c}{${}^{A+1}\text{X}$} & $m(^{A}\text{X}) -
m(^{A+1}\text{X}) + m(^{2}\text{H}) - m(\text{H})$ &
$S_{n}(^{A+1}\text{X}) - S_{n}(^{2}\text{H})$ &
\multicolumn{1}{c}{relative difference} \\
        & (u) & (u) &
        \multicolumn{1}{c}{$(\text{col3}-\text{col2})/\text{col3}
          \times 10^{7}$} \\
\hline                                                  
$^{29}$Si & 0.00670861569(47) & 0.00670860929(536) & -9.54(8.02) \\
\hline                                                  
$^{33}$S & 0.00688901053(50) & 0.00688901206(351) & 2.22(5.15) \\
\hline                                                  
weighted average \\
relative difference & & & -1.21(4.33) \\
\end{tabular}                                                   
\end{ruledtabular}
\end{table*}

The most accurate Si and S gamma-ray energies were measured using
Ge(Li) solid state spectrometers.  These spectrometers derive an
energy scale from gamma-ray standard energies.  For high precision
comparisons, the published values of the gamma-ray energies need to be
adjusted to account for changes in the energy standards.  Because a
number of energy standards covering a wide energy range are used, the
shift in the energy scale for a particular energy is difficult to
estimate.  Although the procedure that has been used is not rigorous,
it is likely sufficient given the accuracy of the published energies.
New values for the standards were taken from Ref.~\onlinecite{helm00}
or have been determined using atomic mass values from
Refs.~\onlinecite{mohr02}, \onlinecite{amdc}, and \onlinecite{audi95}.
For $^{29}$Si the values of the 3539 keV and the 4945 keV energies in
Ref.~\onlinecite{rama92} have been corrected by $-7$ eV to account for
the change in the 2223 keV and the 4945 keV standards produced in the
${}^{1}\text{H}(n,\gamma)$ and ${}^{12}\text{C}(n,\gamma)$ reactions,
respectively.  For $^{33}$S the values for the 841 keV, 2380 keV, and
5421 keV energies in Ref.~\onlinecite{rama85} have been corrected by
$+8$ eV, $-107$ eV, and $-107$ eV respectively.  These corrections
account for changes in the 412 keV standard produced in the decay of
$^{198}$Au and changes in the 2223 keV, the 4945 keV, and the 10829
keV standards produced in the ${}^{1}\text{H}(n,\gamma)$, the
${}^{12}\text{C}(n,\gamma)$, and the ${}^{15}\text{N}(n,\gamma)$
reactions, respectively.  For the low energy Cl lines considerably
more precise data exists.  In 1985 energy values for some Cl lines
were measured using the GAMS4 facility in its early stage of
development \cite{kess85}.  These published values also need to be
corrected for changes in the fundamental constants \cite{mohr02} and
known errors in the lattice spacing of the crystals \cite{desl87}.
These corrections to the Cl lines can be made with much more certainty
than the corrections to the $^{29}$Si and $^{33}$S lines.

\begin{table}
  \caption{\label{tab:comparison}Comparison of measured gamma-ray energies.}
\begin{ruledtabular}
\begin{tabular}{l....}
  \multicolumn{1}{c}{Nuclide} & \multicolumn{1}{c}{This report} &
        \multicolumn{1}{c}{Other references\footnote{$^{29}$Si
        Ref.~\onlinecite{rama92}; $^{33}$S Ref.~\onlinecite{rama85};
        $^{36}$Cl Ref.~\onlinecite{kess85}}} \\
        & \multicolumn{1}{c}{$E_{\text{meas}}$} &
        \multicolumn{1}{c}{$E_{\text{meas}}$}  \\
        & \multicolumn{1}{c}{(eV)} & \multicolumn{1}{c}{(eV)} \\
        \hline
        $^{29}$Si & 3538966.3(1.6) & 3538973(40) \\
        & 4933946.3(4.0) & 4933973(30) \\
        \hline
        $^{33}$S & 840974.08(28) & 840982(14) \\
        & 2379557.6(1.1) & 2379550(11) \\
        & 5420525.5(2.3) & 5420473(40) \\
        \hline
        $^{36}$Cl & 517069.62(22) & 517070.10(23) \\
        & 786296.43(53) & 786297.02(39) \\
        & 1951126.47(89) & 1951127.92(1.37) \\
\end{tabular}
\end{ruledtabular}
\end{table}

In Table~\ref{tab:comparison} we compare the $E_{\text{meas}}$ values
in this report (column 2) with the corrected $E_{\text{meas}}$ values
from other references (column 3).  For the $^{29}$Si and $^{33}$S
gamma-rays the new measurements agree with the corrected older
measurements within the uncertainty except for the 5421 keV line which
differs by 1.3 times the combined uncertainty.  Because of the large
uncertainty of the older Si and S measurements, this comparison does
not provide a very stringent test of our new measurements.  However,
the consistency of the Cl measurements over more than 15 years during
which the spectrometer, the crystals, and measurement procedures were
significantly changed lends a large measure of confidence to the
gamma-ray measurements.

\begin{acknowledgments}
  We thank Albert Henins for preparing the diffracting crystals that
  are used on GAMS4.
\end{acknowledgments}

\bibliography{gams4}

\begin{thebibliography}{34}
\expandafter\ifx\csname natexlab\endcsname\relax\def\natexlab#1{#1}\fi
\expandafter\ifx\csname bibnamefont\endcsname\relax
  \def\bibnamefont#1{#1}\fi
\expandafter\ifx\csname bibfnamefont\endcsname\relax
  \def\bibfnamefont#1{#1}\fi
\expandafter\ifx\csname citenamefont\endcsname\relax
  \def\citenamefont#1{#1}\fi
\expandafter\ifx\csname url\endcsname\relax
  \def\url#1{\texttt{#1}}\fi
\expandafter\ifx\csname urlprefix\endcsname\relax\def\urlprefix{URL }\fi
\providecommand{\bibinfo}[2]{#2}
\providecommand{\eprint}[2][]{\url{#2}}

\bibitem[{\citenamefont{Deslattes and {Kessler, Jr.}}(1979)}]{desl79}
\bibinfo{author}{\bibfnamefont{R.~D.} \bibnamefont{Deslattes}}
  \bibnamefont{and} \bibinfo{author}{\bibfnamefont{E.~G.}
  \bibnamefont{{Kessler, Jr.}}}, in \emph{\bibinfo{booktitle}{Atomic Masses and
  Fundamental Constants-6}}, edited by \bibinfo{editor}{\bibfnamefont{J.~A.}
  \bibnamefont{{Nolan Jr.}}} \bibnamefont{and}
  \bibinfo{editor}{\bibfnamefont{W.}~\bibnamefont{Benenson}}
  (\bibinfo{publisher}{Plenum Press}, \bibinfo{address}{New York},
  \bibinfo{year}{1979}), pp. \bibinfo{pages}{203--218}.

\bibitem[{\citenamefont{Mohr and Taylor}()}]{mohr02}
\bibinfo{author}{\bibfnamefont{P.~J.} \bibnamefont{Mohr}} \bibnamefont{and}
  \bibinfo{author}{\bibfnamefont{B.~N.} \bibnamefont{Taylor}},
  \emph{\bibinfo{title}{The 2002 {CODATA} recommended values of the fundamental
  physical constants, web version 4.0}}, \bibinfo{note}{available at
  physics.nist.gov/constants (National Institute of Standards and Technology,
  Gaithersburg, MD 20899, 9 December 2003).}

\bibitem[{\citenamefont{Lone et~al.}(1981)\citenamefont{Lone, Leavitt, and
  Harrison}}]{lone81}
\bibinfo{author}{\bibfnamefont{M.~A.} \bibnamefont{Lone}},
  \bibinfo{author}{\bibfnamefont{R.~A.} \bibnamefont{Leavitt}},
  \bibnamefont{and} \bibinfo{author}{\bibfnamefont{D.~A.}
  \bibnamefont{Harrison}}, \bibinfo{journal}{Atomic Data and Nuclear Data
  Tables} \textbf{\bibinfo{volume}{26}}, \bibinfo{pages}{511}
  (\bibinfo{year}{1981}).

\bibitem[{\citenamefont{Firestone et~al.}(1996)\citenamefont{Firestone,
  Shirley, Chu, Baglin, and Zipkin}}]{fire96}
\bibinfo{author}{\bibfnamefont{R.~B.} \bibnamefont{Firestone}},
  \bibinfo{author}{\bibfnamefont{V.~S.} \bibnamefont{Shirley}},
  \bibinfo{author}{\bibfnamefont{S.~Y.~F.} \bibnamefont{Chu}},
  \bibinfo{author}{\bibfnamefont{C.~M.} \bibnamefont{Baglin}},
  \bibnamefont{and} \bibinfo{author}{\bibfnamefont{J.}~\bibnamefont{Zipkin}},
  \emph{\bibinfo{title}{Table of Isotopes}} (\bibinfo{publisher}{John Wiley and
  Sons, Inc.}, \bibinfo{address}{New York, New York}, \bibinfo{year}{1996}),
  \bibinfo{edition}{eighth} ed.

\bibitem[{\citenamefont{{Kessler, Jr.} et~al.}(2001)\citenamefont{{Kessler,
  Jr.}, Dewey, Deslattes, Henins, B{\"{o}}rner, Jentschel, and
  Lehmann}}]{kess01}
\bibinfo{author}{\bibfnamefont{E.~G.} \bibnamefont{{Kessler, Jr.}}},
  \bibinfo{author}{\bibfnamefont{M.~S.} \bibnamefont{Dewey}},
  \bibinfo{author}{\bibfnamefont{R.~D.} \bibnamefont{Deslattes}},
  \bibinfo{author}{\bibfnamefont{A.}~\bibnamefont{Henins}},
  \bibinfo{author}{\bibfnamefont{H.~G.} \bibnamefont{B{\"{o}}rner}},
  \bibinfo{author}{\bibfnamefont{M.}~\bibnamefont{Jentschel}},
  \bibnamefont{and} \bibinfo{author}{\bibfnamefont{H.}~\bibnamefont{Lehmann}},
  \bibinfo{journal}{Nucl. Instrum. Methods Phys. Research}
  \textbf{\bibinfo{volume}{A457}}, \bibinfo{pages}{187} (\bibinfo{year}{2001}).

\bibitem[{\citenamefont{{Kessler, Jr.} et~al.}(1999)\citenamefont{{Kessler,
  Jr.}, Dewey, Deslattes, Henins, B{\"{o}}rner, Jentschel, Doll, and
  Lehmann}}]{kess99}
\bibinfo{author}{\bibfnamefont{E.~G.} \bibnamefont{{Kessler, Jr.}}},
  \bibinfo{author}{\bibfnamefont{M.~S.} \bibnamefont{Dewey}},
  \bibinfo{author}{\bibfnamefont{R.~D.} \bibnamefont{Deslattes}},
  \bibinfo{author}{\bibfnamefont{A.}~\bibnamefont{Henins}},
  \bibinfo{author}{\bibfnamefont{H.~G.} \bibnamefont{B{\"{o}}rner}},
  \bibinfo{author}{\bibfnamefont{M.}~\bibnamefont{Jentschel}},
  \bibinfo{author}{\bibfnamefont{C.}~\bibnamefont{Doll}}, \bibnamefont{and}
  \bibinfo{author}{\bibfnamefont{H.}~\bibnamefont{Lehmann}},
  \bibinfo{journal}{Phys. Lett. A} \textbf{\bibinfo{volume}{A255}},
  \bibinfo{pages}{221} (\bibinfo{year}{1999}).

\bibitem[{dis()}]{disclaimer}
\bibinfo{note}{The identification of the supplier of the crystal material is
  included to more completely describe the experiment. Such identification does
  not suggest endorsement nor indicate that this material is necessarily best
  suited for this application.}

\bibitem[{\citenamefont{Becker et~al.}(1981)\citenamefont{Becker, Dorenwendt,
  Ebeling, Lauer, Lucas, Probst, Rademacher, Reim, Seyfried, and
  Siegert}}]{beck81}
\bibinfo{author}{\bibfnamefont{P.}~\bibnamefont{Becker}},
  \bibinfo{author}{\bibfnamefont{K.}~\bibnamefont{Dorenwendt}},
  \bibinfo{author}{\bibfnamefont{G.}~\bibnamefont{Ebeling}},
  \bibinfo{author}{\bibfnamefont{R.}~\bibnamefont{Lauer}},
  \bibinfo{author}{\bibfnamefont{W.}~\bibnamefont{Lucas}},
  \bibinfo{author}{\bibfnamefont{R.}~\bibnamefont{Probst}},
  \bibinfo{author}{\bibfnamefont{H.-J.} \bibnamefont{Rademacher}},
  \bibinfo{author}{\bibfnamefont{G.}~\bibnamefont{Reim}},
  \bibinfo{author}{\bibfnamefont{P.}~\bibnamefont{Seyfried}}, \bibnamefont{and}
  \bibinfo{author}{\bibfnamefont{H.}~\bibnamefont{Siegert}},
  \bibinfo{journal}{Phys. Rev. Lett.} \textbf{\bibinfo{volume}{46}},
  \bibinfo{pages}{1540} (\bibinfo{year}{1981}).

\bibitem[{\citenamefont{Basile et~al.}(1994)\citenamefont{Basile, Bergamin,
  Cavagnero, Mana, Vittone, and Zosi}}]{basi94}
\bibinfo{author}{\bibfnamefont{G.}~\bibnamefont{Basile}},
  \bibinfo{author}{\bibfnamefont{A.}~\bibnamefont{Bergamin}},
  \bibinfo{author}{\bibfnamefont{G.}~\bibnamefont{Cavagnero}},
  \bibinfo{author}{\bibfnamefont{G.}~\bibnamefont{Mana}},
  \bibinfo{author}{\bibfnamefont{E.}~\bibnamefont{Vittone}}, \bibnamefont{and}
  \bibinfo{author}{\bibfnamefont{G.}~\bibnamefont{Zosi}},
  \bibinfo{journal}{Phys. Rev. Lett.} \textbf{\bibinfo{volume}{72}},
  \bibinfo{pages}{3133} (\bibinfo{year}{1994}).

\bibitem[{\citenamefont{Nakayama and Fujimoto}(1997)}]{naka97}
\bibinfo{author}{\bibfnamefont{K.}~\bibnamefont{Nakayama}} \bibnamefont{and}
  \bibinfo{author}{\bibfnamefont{H.}~\bibnamefont{Fujimoto}},
  \bibinfo{journal}{IEEE Trans. Instr. and Meas.}
  \textbf{\bibinfo{volume}{46}}, \bibinfo{pages}{580} (\bibinfo{year}{1997}).

\bibitem[{\citenamefont{Windisch and Becker}(1990)}]{wind90}
\bibinfo{author}{\bibfnamefont{D.}~\bibnamefont{Windisch}} \bibnamefont{and}
  \bibinfo{author}{\bibfnamefont{P.}~\bibnamefont{Becker}},
  \bibinfo{journal}{Phys. Status Solidi} \textbf{\bibinfo{volume}{118}},
  \bibinfo{pages}{379} (\bibinfo{year}{1990}).

\bibitem[{\citenamefont{Kessler et~al.}(1994)\citenamefont{Kessler, Henins,
  Deslattes, Nielsen, and Arif}}]{kess94}
\bibinfo{author}{\bibfnamefont{E.~G.} \bibnamefont{Kessler}},
  \bibinfo{author}{\bibfnamefont{A.}~\bibnamefont{Henins}},
  \bibinfo{author}{\bibfnamefont{R.~D.} \bibnamefont{Deslattes}},
  \bibinfo{author}{\bibfnamefont{L.}~\bibnamefont{Nielsen}}, \bibnamefont{and}
  \bibinfo{author}{\bibfnamefont{M.}~\bibnamefont{Arif}},
  \bibinfo{journal}{Phys. Lett. A} \textbf{\bibinfo{volume}{99}},
  \bibinfo{pages}{1} (\bibinfo{year}{1994}).

\bibitem[{\citenamefont{Kessler et~al.}(1997)\citenamefont{Kessler, Schweppe,
  and Deslattes}}]{kess97}
\bibinfo{author}{\bibfnamefont{E.~G.} \bibnamefont{Kessler}},
  \bibinfo{author}{\bibfnamefont{J.~E.} \bibnamefont{Schweppe}},
  \bibnamefont{and} \bibinfo{author}{\bibfnamefont{R.~D.}
  \bibnamefont{Deslattes}}, \bibinfo{journal}{IEEE Trans. Instr. and Meas.}
  \textbf{\bibinfo{volume}{46}}, \bibinfo{pages}{551} (\bibinfo{year}{1997}).

\bibitem[{\citenamefont{Cavagnero
  et~al.}(2004{\natexlab{a}})\citenamefont{Cavagnero, Fujimoto, Mana, Massa,
  Nakayama, and Zosi}}]{cava03}
\bibinfo{author}{\bibfnamefont{G.}~\bibnamefont{Cavagnero}},
  \bibinfo{author}{\bibfnamefont{H.}~\bibnamefont{Fujimoto}},
  \bibinfo{author}{\bibfnamefont{G.}~\bibnamefont{Mana}},
  \bibinfo{author}{\bibfnamefont{E.}~\bibnamefont{Massa}},
  \bibinfo{author}{\bibfnamefont{K.}~\bibnamefont{Nakayama}}, \bibnamefont{and}
  \bibinfo{author}{\bibfnamefont{G.}~\bibnamefont{Zosi}},
  \bibinfo{journal}{Metrologia} \textbf{\bibinfo{volume}{41}},
  \bibinfo{pages}{56} (\bibinfo{year}{2004}{\natexlab{a}}).

\bibitem[{\citenamefont{Cavagnero
  et~al.}(2004{\natexlab{b}})\citenamefont{Cavagnero, Fujimoto, Mana, Massa,
  Nakayama, and Zosi}}]{cava04}
\bibinfo{author}{\bibfnamefont{G.}~\bibnamefont{Cavagnero}},
  \bibinfo{author}{\bibfnamefont{H.}~\bibnamefont{Fujimoto}},
  \bibinfo{author}{\bibfnamefont{G.}~\bibnamefont{Mana}},
  \bibinfo{author}{\bibfnamefont{E.}~\bibnamefont{Massa}},
  \bibinfo{author}{\bibfnamefont{K.}~\bibnamefont{Nakayama}}, \bibnamefont{and}
  \bibinfo{author}{\bibfnamefont{G.}~\bibnamefont{Zosi}},
  \bibinfo{journal}{Metrologia} \textbf{\bibinfo{volume}{41}},
  \bibinfo{pages}{445} (\bibinfo{year}{2004}{\natexlab{b}}).

\bibitem[{pri({\natexlab{a}})}]{private-taylor}
\bibinfo{note}{Private communication, Barry Taylor, NIST}.

\bibitem[{\citenamefont{Mohr and Taylor}(2000)}]{mohr00}
\bibinfo{author}{\bibfnamefont{P.~J.} \bibnamefont{Mohr}} \bibnamefont{and}
  \bibinfo{author}{\bibfnamefont{B.~N.} \bibnamefont{Taylor}},
  \bibinfo{journal}{Rev. Mod. Phys.} \textbf{\bibinfo{volume}{72}},
  \bibinfo{pages}{351} (\bibinfo{year}{2000}).

\bibitem[{\citenamefont{Birch and Downs}(1993)}]{birc93}
\bibinfo{author}{\bibfnamefont{K.~P.} \bibnamefont{Birch}} \bibnamefont{and}
  \bibinfo{author}{\bibfnamefont{M.~J.} \bibnamefont{Downs}},
  \bibinfo{journal}{Metrologia} \textbf{\bibinfo{volume}{30}},
  \bibinfo{pages}{155} (\bibinfo{year}{1993}), \bibinfo{note}{index of
  refraction stuff}.

\bibitem[{\citenamefont{Birch and Downs}(1994)}]{birc94}
\bibinfo{author}{\bibfnamefont{K.~P.} \bibnamefont{Birch}} \bibnamefont{and}
  \bibinfo{author}{\bibfnamefont{M.~J.} \bibnamefont{Downs}},
  \bibinfo{journal}{Metrologia} \textbf{\bibinfo{volume}{31}},
  \bibinfo{pages}{315} (\bibinfo{year}{1994}), \bibinfo{note}{index of
  refraction stuff}.

\bibitem[{\citenamefont{B{\"{o}}rner and Jolie}(1993)}]{born93}
\bibinfo{author}{\bibfnamefont{H.~G.} \bibnamefont{B{\"{o}}rner}}
  \bibnamefont{and} \bibinfo{author}{\bibfnamefont{J.}~\bibnamefont{Jolie}},
  \bibinfo{journal}{J. Phys. G: Nucl. Phys.} \textbf{\bibinfo{volume}{19}},
  \bibinfo{pages}{217} (\bibinfo{year}{1993}).

\bibitem[{\citenamefont{Bergamin et~al.}(1997)\citenamefont{Bergamin,
  Cavagnero, Mana, and Zosi}}]{berg97}
\bibinfo{author}{\bibfnamefont{A.}~\bibnamefont{Bergamin}},
  \bibinfo{author}{\bibfnamefont{G.}~\bibnamefont{Cavagnero}},
  \bibinfo{author}{\bibfnamefont{G.}~\bibnamefont{Mana}}, \bibnamefont{and}
  \bibinfo{author}{\bibfnamefont{G.}~\bibnamefont{Zosi}}, \bibinfo{journal}{J.
  Appl. Phys.} \textbf{\bibinfo{volume}{82}}, \bibinfo{pages}{5396}
  (\bibinfo{year}{1997}).

\bibitem[{\citenamefont{Berthold and Jacobs}(1976)}]{bert76}
\bibinfo{author}{\bibfnamefont{J.~W.} \bibnamefont{Berthold}} \bibnamefont{and}
  \bibinfo{author}{\bibfnamefont{S.~F.} \bibnamefont{Jacobs}},
  \bibinfo{journal}{Applied Optics} \textbf{\bibinfo{volume}{15}},
  \bibinfo{pages}{1898} (\bibinfo{year}{1976}).

\bibitem[{\citenamefont{Schnopper}(1965)}]{schn65}
\bibinfo{author}{\bibfnamefont{H.~W.} \bibnamefont{Schnopper}},
  \bibinfo{journal}{J. Appl. Phys.} \textbf{\bibinfo{volume}{36}},
  \bibinfo{pages}{1415} (\bibinfo{year}{1965}).

\bibitem[{\citenamefont{Nye}(1957)}]{nye57}
\bibinfo{author}{\bibfnamefont{J.~F.} \bibnamefont{Nye}},
  \emph{\bibinfo{title}{Physical Properties of Crystals}}
  (\bibinfo{publisher}{Oxford University Press}, \bibinfo{address}{Oxford,
  England}, \bibinfo{year}{1957}), pp. \bibinfo{pages}{146--147}.

\bibitem[{\citenamefont{McSkimin}(1953)}]{mcsk53}
\bibinfo{author}{\bibfnamefont{H.~J.} \bibnamefont{McSkimin}},
  \bibinfo{journal}{J. Appl. Phys.} \textbf{\bibinfo{volume}{24}},
  \bibinfo{pages}{988} (\bibinfo{year}{1953}).

\bibitem[{\citenamefont{Ko et~al.}(1999)\citenamefont{Ko, Cheoun, and
  Cheon}}]{yong99}
\bibinfo{author}{\bibfnamefont{Y.}~\bibnamefont{Ko}},
  \bibinfo{author}{\bibfnamefont{M.~K.} \bibnamefont{Cheoun}},
  \bibnamefont{and} \bibinfo{author}{\bibfnamefont{I.-T.} \bibnamefont{Cheon}},
  \bibinfo{journal}{Phys. Rev. C} \textbf{\bibinfo{volume}{59}},
  \bibinfo{pages}{3473} (\bibinfo{year}{1999}).

\bibitem[{amd()}]{amdc}
\emph{\bibinfo{title}{The atomic mass data center}},
  \bibinfo{note}{\url{http://www.nndc.bnl.gov/amdc/}}.

\bibitem[{\citenamefont{Audi and Wapstra}(1995)}]{audi95}
\bibinfo{author}{\bibfnamefont{G.}~\bibnamefont{Audi}} \bibnamefont{and}
  \bibinfo{author}{\bibfnamefont{A.~H.} \bibnamefont{Wapstra}},
  \bibinfo{journal}{Nucl. Phys. A} \textbf{\bibinfo{volume}{595}},
  \bibinfo{pages}{409} (\bibinfo{year}{1995}).

\bibitem[{pri({\natexlab{b}})}]{private-pritchard}
\bibinfo{note}{Private communication, Professor David E. Pritchard, MIT}.

\bibitem[{\citenamefont{Helmer and van~der Leun}(2000)}]{helm00}
\bibinfo{author}{\bibfnamefont{R.~G.} \bibnamefont{Helmer}} \bibnamefont{and}
  \bibinfo{author}{\bibfnamefont{C.}~\bibnamefont{van~der Leun}},
  \bibinfo{journal}{Nucl. Instrum. Methods A} \textbf{\bibinfo{volume}{450}},
  \bibinfo{pages}{35} (\bibinfo{year}{2000}).

\bibitem[{\citenamefont{Raman et~al.}(1992)\citenamefont{Raman, Jurney,
  Starner, and Lynn}}]{rama92}
\bibinfo{author}{\bibfnamefont{S.}~\bibnamefont{Raman}},
  \bibinfo{author}{\bibfnamefont{E.~T.} \bibnamefont{Jurney}},
  \bibinfo{author}{\bibfnamefont{J.~W.} \bibnamefont{Starner}},
  \bibnamefont{and} \bibinfo{author}{\bibfnamefont{J.~E.} \bibnamefont{Lynn}},
  \bibinfo{journal}{Phys. Rev. C} \textbf{\bibinfo{volume}{46}},
  \bibinfo{pages}{972} (\bibinfo{year}{1992}).

\bibitem[{\citenamefont{Raman et~al.}(1985)\citenamefont{Raman, Carlton, Wells,
  Jurney, and Lynn}}]{rama85}
\bibinfo{author}{\bibfnamefont{S.}~\bibnamefont{Raman}},
  \bibinfo{author}{\bibfnamefont{R.~F.} \bibnamefont{Carlton}},
  \bibinfo{author}{\bibfnamefont{J.~C.} \bibnamefont{Wells}},
  \bibinfo{author}{\bibfnamefont{E.~T.} \bibnamefont{Jurney}},
  \bibnamefont{and} \bibinfo{author}{\bibfnamefont{J.~E.} \bibnamefont{Lynn}},
  \bibinfo{journal}{Phys. Rev. C} \textbf{\bibinfo{volume}{32}},
  \bibinfo{pages}{18} (\bibinfo{year}{1985}).

\bibitem[{\citenamefont{{Kessler, Jr.} et~al.}(1985)\citenamefont{{Kessler,
  Jr.}, Greene, Deslattes, and B{\"{o}}rner}}]{kess85}
\bibinfo{author}{\bibfnamefont{E.~G.} \bibnamefont{{Kessler, Jr.}}},
  \bibinfo{author}{\bibfnamefont{G.~L.} \bibnamefont{Greene}},
  \bibinfo{author}{\bibfnamefont{R.~D.} \bibnamefont{Deslattes}},
  \bibnamefont{and} \bibinfo{author}{\bibfnamefont{H.~G.}
  \bibnamefont{B{\"{o}}rner}}, \bibinfo{journal}{Phys. Rev. C}
  \textbf{\bibinfo{volume}{32}}, \bibinfo{pages}{374} (\bibinfo{year}{1985}).

\bibitem[{\citenamefont{Deslattes et~al.}(1987)\citenamefont{Deslattes, Tanaka,
  Greene, Henins, and {Kessler, Jr.}}}]{desl87}
\bibinfo{author}{\bibfnamefont{R.~D.} \bibnamefont{Deslattes}},
  \bibinfo{author}{\bibfnamefont{M.}~\bibnamefont{Tanaka}},
  \bibinfo{author}{\bibfnamefont{G.~L.} \bibnamefont{Greene}},
  \bibinfo{author}{\bibfnamefont{A.}~\bibnamefont{Henins}}, \bibnamefont{and}
  \bibinfo{author}{\bibfnamefont{E.~G.} \bibnamefont{{Kessler, Jr.}}},
  \bibinfo{journal}{IEEE Trans. Instr. and Meas.}
  \textbf{\bibinfo{volume}{36}}, \bibinfo{pages}{1660169}
  (\bibinfo{year}{1987}).

\end{thebibliography}

\end{document}